\documentstyle[12pt]{article}
\begin{document}

\centerline {\Large \bf Self-Organized Percolation Model}

\centerline {\Large \bf for Stock Market Fluctuations}

\bigskip
\centerline {Dietrich Stauffer$^1$ and Didier Sornette$^{2,3}$}

\noindent
$^1$ Institute for Theoretical Physics, Cologne University, 50923 K\"oln,
Germany
\centerline {e-mail: stauffer@thp.uni-koeln.de}

\noindent
$^2$ Laboratoire de Physique de la Mati\`ere Condens\'ee, CNRS UMR6622 and
Universit\'e de Nice-Sophia Antipolis, Facult\'e des Sciences, B.P. 71
06108 NICE Cedex 2, France

\noindent
$^3$  Institute of Geophysics and Planetary Physics
and Department of Earth and Space Science,
University of California, Los Angeles, California 90095
\centerline {e-mail: sornette@cyclop.ess.ucla.edu}

\bigskip
\bigskip
ABSTRACT: In the Cont-Bouchaud model [cond-mat/9712318] of stock markets,
percolation clusters act as buying or selling investors and their
statistics controls that of the price variations. Rather than fixing
the concentration controlling each cluster connectivity artificially
at or close to the critical value, we propose that
clusters shatter and aggregate continuously as the
concentration evolves randomly, reflecting
the incessant time evolution of groups of opinions and market moods.
By the mechanism of ``sweeping of an instability''
[D. Sornette, Journal de Physique I 4, 209 (1994)], this market model
spontaneously exhibits reasonable power law
statistics for the distribution of price changes and accounts for the other
important stylized facts of stock market price fluctuations.


\bigskip \bigskip

\section{The Percolation model of stock market prices}

A wealth of models \cite{Lo,lux} (to our knowledge, the first
stock market simulation was performed by the economist Stigler in 1964
\cite{Stigler}),
partially listed in \cite{sob}, have been introduced in the
financial and more recently in the physical community which attempt to capture
the complex behavior of stock market prices and of market participants.
Based on the competition between supply and demand, the effort is
to model the main observed stylized facts: absence of two-point correlation of
the returns \cite{Lo}, fat tail distribution of returns
(probabilities higher than Gaussian \cite{lux,Lahsor,bpbook})
and long-range volatility (standard deviation) correlations
\cite{Granger}. The goal is to have, on the one hand, the simplest and most
parsimonious
description of the market and, on the other hand, the most faithful
representation
of the observed market characteristics.

In this spirit, Cont and Bouchaud \cite{cb} introduced a percolation model
which assumes that investors can
be classified into groups (clusters) of the same opinion occurring with
many different
sizes. The simplest
recipe to aggregate interacting or inter-influencing traders into
groups is to assume that the connectivity
between traders defining the groups can be seen as a pure
geometrical
percolation problem with fixed occupancy on a given network topology.
Clusters are groups of neighboring occupied sites or investors.
Then, random percolation clusters make
a decision to buy or sell on the stock market, for all sites (corresponding
to the individual investors and units of wealth) in  that cluster
together. Thus, the individual investors are thought to cluster together to
form companies or groups of influence, which under the guidance of a single
manager buy
(probability $a$), sell (probability $a$), or refrain from trading (probability
$1-2a$) within one time interval. The traded amount is proportional to the
number $s$ of sites in the cluster, and
$$
{\rm the ~logarithm~of~the~price~changes ~proportionally ~to}
$$
\begin{equation}
{\rm the ~difference ~\Delta~ between ~demand ~and ~supply~.}
\label{eakjka}
\end{equation}

When the activity $a$
is small, at most one cluster trades at a time. As a consequence,
the distribution $P(R)$ of relative price changes or ``returns''
$R$ scales as the well-known \cite{perc} cluster size distribution $n_s(p)$
of percolation theory. In contrast, for large activity $a$
and without an infinite cluster, the relative
price variation is the contribution (sum) of many clusters and the central
limit
theorem implies that the distribution $P(R)$ converges to the Gaussian law for
large systems (except exactly at the critical point $p_c$).

For low activity $a$ and right at the site percolation threshold $p=p_c$, when
the fraction $p$ of lattice sites occupied by an investor
in a $d$-dimensional lattice of linear extent $L$ barely suffices to form an
``infinite'' cluster stretching from top to bottom, we observe power laws:
\begin{equation}
n_s \propto s^{-\tau}, ~~~~~ P(R) \propto R^{-\tau}~~~{\rm for}~~1 \ll s
\ll L^D
~~~~{\rm where}~ D = d/(\tau-1)   \label{ytfvqv}
\end{equation}
is the fractal dimension of the percolating cluster.

For concentrations $p$ different from $p_c$, the cluster numbers decay as an
exponential (resp.  stretched exponential)
for $p < p_c$ (resp. for $p > p_c$). Their typical size $s^*$ (not counting
the ``infinite'' percolating cluster) is much smaller than
at $p = p_c$ where it is solely controlled by the system size (i.e.
total number of traders).  Since a cluster of size roughly comparable to the
total system size appears at and above the percolation threshold $p_c$, this
value corresponds to a big crash in the market, and
the region of $p$ below $p_c$ gives less volatile behavior.

Correlations of volatility in time are produced \cite{sob} by letting the
occupied sites (traders)
diffuse slowly to empty neighbor sites on a lattice to reform or destroy new
alliances.  The volatility,
i.e. the typical absolute value of the return, thus behaves similar to a mean
cluster size and is correlated in time due to the slow diffusion \cite{sob}
with a typical decay slower than exponential.

This model at the percolation threshold thus agrees qualitatively (but
not quantitatively especially on the exponents as discussed below) with the
three stylized
facts of real markets \cite{Lo,lux}: the average return $R$ is zero (if
inflation
and other regular trends are subtracted); the return distribution $P(R)$ decays
as a power law $\propto R^{-\mu}$ for intermediate $R$ with $\mu \simeq 4$
(the probability
$P_>(R)$ of finding a change larger than $R$ then varies as $R^{1-\mu}$),
and the volatility
\begin{equation}
V(t)= <R(t)^2>^{1/2}
\end{equation}
clusters in time in the sense that its autocorrelation
function
\begin{equation}
C(\tau) =  <\Delta V(t) \cdot \Delta V(t+\tau)> ~,~~~~{\rm with} ~~\Delta V
= V - <V>~,
\end{equation}
is positive and decays slowly to zero.

Two disadvantages of the model are:
\begin{itemize}
\item why should markets know the percolation threshold and work at $p=p_c$?
\item How can a value of $\mu = \tau$ be nearly
$4$ when $\tau$ varies only from 2 to 2.5 if the dimensionality $d$
increases from $2$ to
infinity ?
\end{itemize}
A mechanism for self-organized criticality like invasion
percolation \cite{perc} would only solve the first and not the second problem
and would be difficult to justify form an economic view point.

\section{A Simple Self-Organizing Market}

\subsection{Percolation connectivity evolving with time}

We thus return to an alternative mechanism \cite{sornette} which gives power
laws without the need to tune $p$ to $p_c$ and which is very robust and simple.
The new idea we propose in this context is that there is no reason a
priori to expect that the
parameter $p$ controlling the connectivity/influence between traders is fixed.
The circle of professionals and colleagues to whom a trader is typically
connected
evolves as a function of time, not only in its structure at fixed average
number
of connections (corresponding to the
diffusion effect discussed above) but also, in the average strength
and number of interactions: at some times, traders are following strong
herding
behavior and the effective connectivity parameter $p$ is high; at other times,
investors are more individualistic and smaller values of $p$ seem more
reasonable.
In order to take into account the complex dynamics of the network of
interactions
between traders, it thus seems reasonable to relax the hypothesis that $p$
is fixed
at a given value but rather evolves with its own dynamics. The simplest
version is
to assume that $p$ is taken purely random at each time step. As a consequence,
the distribution of relative price changes will be an average over those
obtained
for each sampled $p$'s. Averaging \cite{sornette} over an interval in $p$
containing the percolation threshold $p_c$ will give its main contribution
to the number of large clusters from a narrow region (width $\propto
1/s^\sigma$) about $p_c$, and thus lead to an integrated cluster numbers
$\propto
s^{-\tau - \sigma}$, where $\sigma$ varies from $0.4$ to $0.5$ if $d$
varies from
$2$ to infinity. Now $\mu = \tau + \sigma$ varies from $2.45$ to $3$, closer to
reality ($\mu \simeq 4$).

We stress that, as soon as $p$ samples an interval containing or close to
$p_c$,
the distribution of returns is a power law with no other truncation than
given by the finite size of the total system. We thus obtain a robust critical
behavior without any artificial adjustement of the connectivity parameter $p$.
Notice that this mechanism in terms of a ``sweeping of an instability''
\cite{sornette}
is different from what is usually called self-organized criticality \cite{bak}
which involves a dynamical feedback attracting the system dynamics to a
dynamical
critical point.

We thus \cite{sob} distribute randomly our investors on the $L \times L$
square lattice, with concentration $p$. We sum up all the results obtained by
 varying $p$ in steps of one percent, from 1 to 59 percent where the
percolation threshold $p_c = 0.593$ is reached. For each concentration, we
make 1000 iterations where at each iteration one percent of the investors try
to move to a randomly selected neighbor site. For each cluster configuration
obtained in this way, we sum over 1000 different realizations of buying and
selling decisions of the cluster, which
thus allows much better averaging than in real markets where history cannot be
repeated so easily.  Many such simulations are averaged over to give smooth
results.

Fama has argued \cite{Fama} that the crash of Oct. 1987 on the US
and other stock markets worldwide could be seen as the signature of an
efficient
reassessement of and convergence to the correct ``fundamental'' price after
the long
speculative
bubble preceeding it. In this spirit, we assume that, as $p$
reaches the crash concentration $p=p_c$,
``everything'' changes and the network connectivity is afterwards
reinitialized at a value $p< p_c$; therefore no data for $p > p_c$  are
given here.

Fig.1 shows for the square lattice ($d=2$) the distribution of returns
$P(R)$. The
simulations confirm, for a range of about five orders of magnitude in $P$
similar to the range observed by Gopikrishnan et al \cite{lux},
the predicted power law $P(R) \propto 1/R^\mu$ at intermediate $R$ with an
exponent
$\mu \simeq 2.5$. For the largest $R$, finite size effects
reduce $P(R)$ and, for small $R$, the probability is roughly constant.
Increasing
the linear lattice size $L$ from 31 via 101 to 301 shifts the power-law
region to
larger $R$ without changing the effective exponent.

\subsection{Size-dependent activity}

The numerical deviation from the empirically observed $\mu = 4$ is thus
large, and a different approach is needed. Instead of
taking the activity $a$ as a free parameter between zero and $1/2$ ($0.005$
in Fig.1)
and the same independently of the cluster size, we assume the following size
dependence
\begin{equation}
a = 0.5/\sqrt{s}~,  \label{jajak}
\end{equation}
thus getting rid of one free parameter. A priori, it is reasonable to consider
that the big investors, such as the mutual and/or retirement funds with
their prudent approach, their emphasis on
low risk, and their enormous inertia due to the fact that large positions
move the market unfavorably, have to and do trade less often than small
professional
investors who have to generate their income from active trading
rather than from sheer mass. In this spirit, recent works have documented that
the growth dynamics of business firms \cite{firms}, the economies of countries
\cite{countries} and the university
research activities \cite{univ} depend on size, the smaller entities being
the most active
proportionally. Another not necessarily exclusive mechanism is that, within
a large cluster, the $s$ investors have to agree by some majority
to buy and sell, and do not trade if no such majority is reached. A random
decision process then could lead to the square-root behavior given by
(\ref{jajak}).

With this modification, the exponent $\mu$ is predicted to be
\begin{equation}
\mu = \tau +  \sigma + 1/2 \simeq 3~. \label{ahaj}
\end{equation}
In contrast, we measure in Fig.2 an effective exponent in the intermediate $R$
range equal to $3.5$, larger than the asymptotically expected value given by
eq.(\ref{ahaj})
and close to the empirical value near $4$. The volatility
clustering is not destroyed by our change, and Fig.3 shows the volatility
auto-correlations to decay slowly towards zero, also in agreement with
empirical facts.

Since eq.(4) implies a zero probability
of the ``infinite'' cluster to act for an ``infinite'' system above $p_c$,
 we now can also integrate over the whole
interval of $p$ by summing from 1 to 99 percent.
Fig.4 documents a new phenomenon, namely
wings with rare price changes
determined mainly by large clusters containing nearly the
whole lattice and choosing randomly to buy, sell, or sleep.
The larger the lattice is, the smaller is the overall
{\it absolute} weight of these wings. However, relative to the power law
represented by the dashed straight line, the larger the lattice, the larger
is the {\it relative} weight of the wings. This means that these large price
changes are more and more ``outliers'' of the power law statistics
holding for intermediate price changes.

To understand this observation, recall first that the connectivity parameter
$p$ goes from a small number ($1\%$ in the simulations) to a larger number
($99\%$ in the simulations) above $p_c$ and the observed distribution of
log-price changes is mapped one-to-one onto the distribution $P_{\rm sum}$
of cluster sizes
obtained by averaging over all the cluster size distributions obtained for
$p$ taken
with equal weight between $1\%$ and $99\%$. If the $n_s$ of eq.(2) are the
cluster numbers per lattice site, then $L^2n_s$ are those in the whole lattice
and vary at $p=p_c$ as $L^2s^{-\tau}$. The contribution $P_{sum}(s)$ of
$s$-clusters averaged over all $p$, to the price changes reads
\begin{equation}
P_{\rm sum} \propto L^2 ~ {1 \over s^{\tau + \sigma + 1/2}}~.  \label{uyashnak}
\end{equation}
The two additional terms $\sigma + 1/2$ in the exponent of (\ref{uyashnak})
stem, as discussed above, from the two
effects of ``sweeping'' of $p$ over $p_c$ and of the
inverse square root dependence of the activity on the cluster size.

Now, extrapolating (\ref{uyashnak}), we can estimate the probability
$cL^2 P_{sum}(cL^2)$ that
this power law
{\it would} predict for getting a cluster of size of order $L^2$ and
compare it to the
{\it true} probability of getting a cluster of this size.
The factor $cL^2$ comes from the fact that one must count the large
clusters around a
typical size with proportional fluctuations, thus giving a true probability
while
$P_{sum}(cL^2)$ is the probability density to observe a cluster of size
$cL^2$. From
(\ref{uyashnak}), we get
\begin{equation}
cL^2 P_{sum}(cL^2) \propto (cL^2) ~L^2 ~ {1 \over (cL^2)^{\tau + \sigma +
1/2}} ~,  \label{jhfdjaka}
\end{equation}
where $c$ is a number of order unity.
Counting powers of $L$ in (\ref{jhfdjaka}) and expressing the result for
two dimensions gives
\begin{equation}
cL^2 P_{sum} \propto L^{4-2\tau - 2 (1/2 + \sigma)} = {1 \over L^{1.9}}~.
\end{equation}

In contrast, we know that there is exactly one
infinite cluster in a quadratic lattice above $p_c$. Since the $p$'s are
uniformely taken
between $1\%$ and $99\%$, this shows directly that the probability to get a
cluster close
to the maximum size $L^2$ is a constant fraction,
{\it independently} of the lattice size $L$, and its contribution to a
price change is multiplied by its activity $\propto 1/L$.
(Indeed our data of Fig.4 give a value near $0.5/L$ for the
fraction of returns larger than $L^2/2$.)
This argument thus shows that the
ratio of the true frequency to observe the large ``outlier'' (stemming from
the infinite cluster truncated to the size of the lattice above $p_c$) is
larger than the
extrapolation of the power distribution of intermediate cluster size by a
factor $L^{0.9}$ which increases with the system size $L$, in qualitative
agreement with Fig.4.

We suggest that the large wings might
correspond to the ``outliers'' in the stock
market like the Wall Street crashes of 1929 and 1987 \cite{outliers}.
 The normal autocorrelation
functions of the volatility exhibiting long range dependence then apply to
normal times on the stock market
when no such outliers are relevant, and they are destroyed by the outliers.
 The wings vanish and the
autocorrelations are restored if we follow \cite{sob} and omit the largest
cluster from the market.

\subsection{Nonlinear price change dependence}

All our previous results derive from the assumption (\ref{eakjka})
that the change of (the logarithm of the) price is proportional to the
difference between supply and demand. This assumption is often made
and can be in fact derived rigorously \cite{Farmer} from the two
assumptions that it is not
possible to make profits by repeatedly trading through a circuit and that
the ratio of prices before and after a transaction is a function of the
difference $\Delta$ between demand and supply alone.

However, many recent empirical studies suggest that the relationship
between the change
of the logarithm of price and $\Delta$ is highly nonlinear, especially for
large
orders \cite{Jensen}. Assuming that the time needed to complete a trade of
size $s$
is proportional to $s$ and that the unobservable price fluctuations obey a
diffusion
process during that time, Zhang derives the relationship that the
change of the logarithm of the price is proportional to the square root of
the difference $\Delta$ between demand and supply \cite{Zhang}, i.e. to the
square root of $s$ in
our present formulation. This modifies all previous results as follows.

The result (\ref{ytfvqv}) for the ``pure'' percolation model becomes
\begin{equation}
n_s \propto s^{-\tau}, ~~~~~ P(R) \propto R^{-\mu}~~~{\rm for}~~1 \ll s
\ll L^D ~~~~{\rm where}~ R \propto \sqrt s~,
\end{equation}
giving with $ds/dR \propto R$ (with numerical estimates in two dimensions)
the exponent $\mu = 2\tau -1$ around $3.1$, still smaller than the empirical
value  close to $4$.

The result obtained by the ``sweeping'' of the connectivity parameter $p$
transforms $\mu$ from $\mu = \tau + \sigma$ into $\mu = 2\tau - 1 + \sigma$,
giving a value $3.5$. Next, incorporating the size dependence (\ref{jajak})
of the activity leads to the prediction $\mu = 2\tau + \sigma = 4.5$, in
rough agreement with the empirical value 4.

We may even omit the size-dependent activity and use only this nonlinear
price change dependence and $0 < p < p_c$. Then the data of Fig.1 are
transformed, without any additional simulations, into those of Fig.5 which
give an effective $\mu \simeq 3.9$ in better agreement with the real $\mu
\simeq
4$ than the theoretical prediction $\mu = 2\tau -1 + \sigma \simeq 3.5.$

\section{Concluding Remarks}

We have presented what we believe is probably the simplest and most robust
model of stock market
dynamics without tunable parameters that self-organizes into a regime where
the most important empirical characteristics of stock market price dynamics are
captured.

In this simplest version, we have chosen the most straightforward dynamics
of the interaction/connectivity parameter $p$, i.e. a
continuous increase up to the critical value $p_c$
followed by a reset to a low value and so on. Incorporating a size-dependence
of the cluster activities has allowed us to let $p$ larger than $p_c$ for which
we have documented the appearance of outliers corresponding to the infinite
cluster truncated to the size of the lattice. This outlier might correspond
to the large crashes observed in this century. A good agreement with
empirical data is obtained alternatively by allowing for a nonlinear
dependence of the change of (the logarithm of the) price as a function of the
difference between supply and demand.

A random evolution of $p$,
either pure white noise,
or a random walk or with more correlation are interesting to investigate in
the future, but
will not change the most fundamental finding presented here of a power law
distribution and
long-range correlations of the volatility. More interesting
presumably would be a dynamic of $p$ coupled to that of the price
change, simulating the tendency to join a bullish market \cite{lux}.
Note also that the present model is by construction up-down symmetric,
which means that rallies
appear as often statistically and in the same shape as crashes. There
is not sharp peak versus flat trough asymmetry \cite{RoehnerSor}.
Such asymmetry can be easily incorporated by letting
the trading activity be dependent on the function price(time), i.e.
increasing prices
causes more people to act than a decreasing price, but we have not persued this
as this would imply adding novel ingredients in a model we have on purpose
kept bare to its skeleton.

\vskip 0.5cm
Ackowledgements: This idea originated at the meeting ``Facets of
Universality in Complex Systems:
Climate, Biodynamics and Stock Markets'', organized at Giessen University
by Armin Bunde and John Schellnhuber (June 1999).
We thank NIC J\"ulich for time on their Cray-T3E, T.Lux for ref.3, and A.
Johansen for comments. One of us (DS) wishes to
point out that all errors are due to the other author (DS).

\newpage
Captions:
\parindent 0pt
\vskip 1cm
Fig.1: Return distribution at constant activity $a=0.005$. The axis of relative
price variations is scaled
such that a buying cluster of $s$ investors produces an increase of the price
by $s$.
\vskip 1cm

Fig.2: Return distribution with activity decaying as $1/\sqrt{s}$.

\vskip 1cm
Fig.3: Volatility autocorrelation function $C(T)$ versus time lag $T$ for the
simulations of Fig.2.

\vskip 1cm
Fig.4: As Fig.2, but with the cluster connectivity parameter $p$ varying
from $1$ to $99$ percent.

\vskip 1cm
Fig.5: Data from Fig.1 replotted by assuming a price change
proportional to the square root of the (absolute
value of the) difference $\Delta$ between demand and supply (and with sign
opposite to
that of $\Delta$).

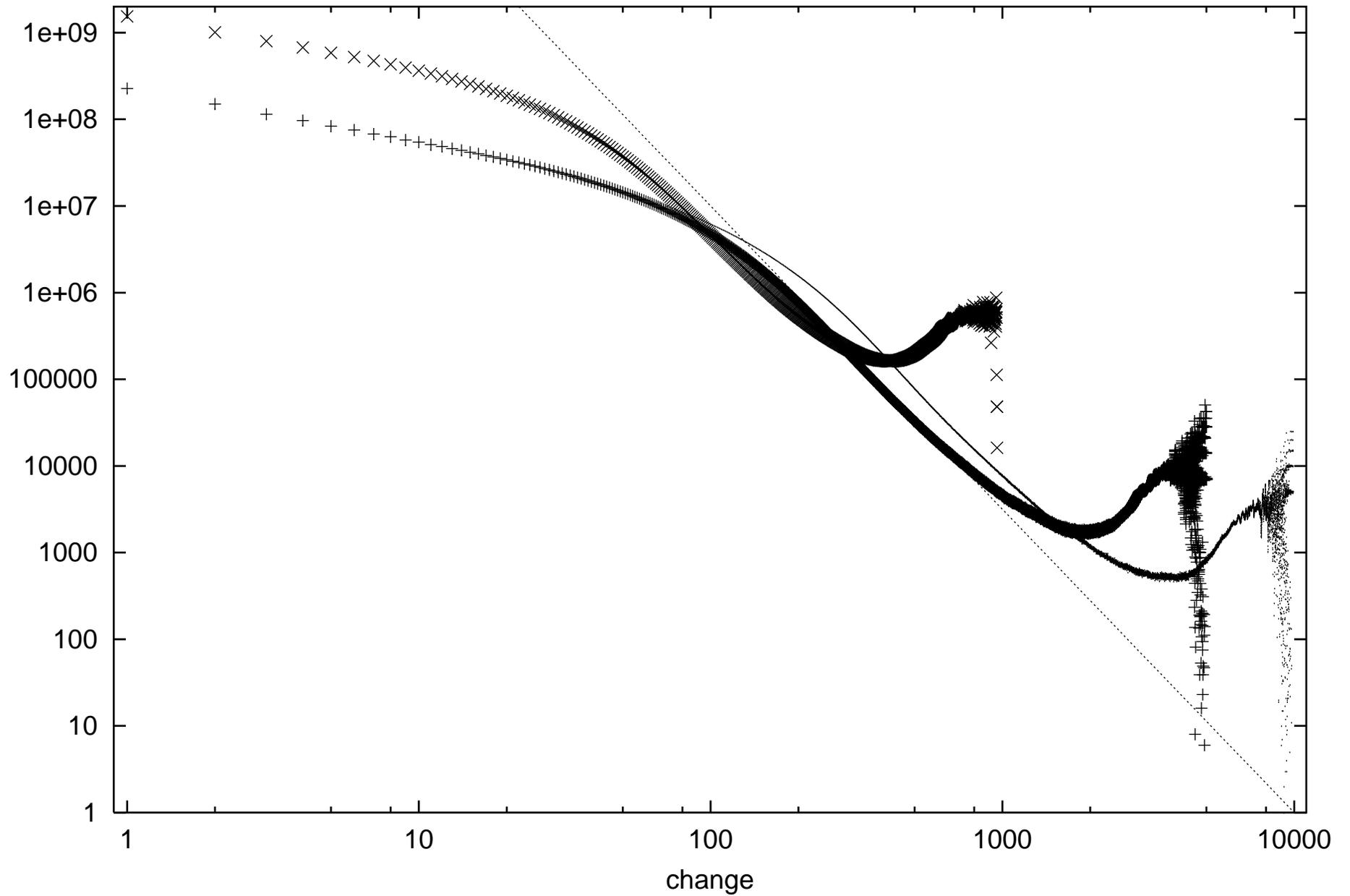
\end{document}